\documentclass[twocolumn,aps]{revtex4}

\newcommand{\ba}{\begin{eqnarray}}
\newcommand{\ea}{\end{eqnarray}}
\newcommand{\be}{\begin{equation}}
\newcommand{\ee}{\end{equation}}
\newcommand{\bib}{\bibitem}
\newcommand{\ed}{\end{document}}

\begin{document}

\title{Testing quantum duality using cold Rydberg atoms}
\vspace{1.0cm}
\author{J.L. Noronha}
\affiliation{Centro Brasileiro de Pesquisas F\'\i sicas (CBPF) 22290-180, Rio de Janeiro, Brazil\\}
\author{C. Wotzasek}
\affiliation{Instituto de F\'\i sica, Universidade
Federal do Rio de Janeiro\\21945, Rio de Janeiro, Brazil\\}

\begin{abstract}
\noindent 
\hspace{3ex}Classically, the dynamics of the chiral oscillator (CO) may be described by the Landau model (LM) through a well established mathematical procedure known as duality mapping. In this letter we show how this duality is broken in quantum mechanics due to the presence of a $Z_2$-anomaly in the CO. We give the theoretical basis for an experimental setup displaying the possibility to measure this global anomaly using cold Rydberg atoms.

\end{abstract}
\maketitle

Duality is a mapping between two different mathematical descriptions of a same physical phenomenon. A strongly (weakly) coupled theory is mapped, through duality, in a weakly (strongly) coupled theory. Therefore, dualities have a striking importance in the study of strongly interacting models. Examples in condensed matter and high energy physics are the Kramers-Wanier high/low temperature mapping, the 4D electromagnetic Montonen-Olive particle/soliton conjecture and the string dualities \cite{reviewsduality}. All these examples involve theories with many degrees of freedom but of course this is not a necessary condition.

In this work we study duality in an example involving finite degrees of freedom, the LM/CO duality. Both models describe the classical trajectories of a particle in the x-y plane subject to a constant magnetic field $\vec{B}=B\widehat{z}$ \cite{landau, mqchernsimons, quiraloscillator}, and therefore, the classical duality would be already expected. The issue of duality at the quantum level is however more subtle. It was shown in \cite{mqchernsimons, maslov} that the CO has a $Z_2$-anomaly that forces its orbital angular momentum to be quantized in half-integer units of $\hbar$, either positive or negative, letting it with an uncommon behavior through rotations. This anomaly has a great resemblance with the $Z_2$-anomaly of a single Weyl fermion coupled to a SU(2) Yang-Mills field discovered by Witten \cite{wittenanomaly}. We show in this Letter that, at the quantum domain, the partition function for the LM can be factorized as the product of two $Z_2$-anomalous theories -- the CO and a ``mechanical Chern-Simons (CS)" model. We discuss how the CO and the LM differently respond to this anomaly and how this process is related with the violation of duality in quantum mechanics. We also discuss the theoretical basis for an interferometer experiment, involving cold Rydberg atoms pointing, at least in principle, a possible direction leading to direct consequence of the $Z_2$ anomaly - the gain of a $\pi$ phase factor by the CO fundamental state - with consequences to the existence (or not) of the quantum duality. 

Now, we analyze the LM/CO duality in a classical and quantum context. The Lagrangian of the LM can be written as
\be
\ L_{LM}= \frac{m\dot{x}_{j}\dot{x}_{j}}{2}-\frac{g}{2}\epsilon_{ij}{x}_{i}\dot{x}_{j}.
\label{LM}
\ee   
where $\epsilon_{ij}$ is the Levi-Civita symbol. The CO model, on the other hand, is given by
\be
\ L_{CO}= \frac{g}{2}\epsilon_{ij}x_{i}\dot{x}_{j}-\frac{k}{2}x_{i}x_{i}, 
\label{CO}
\ee
where we consider $g$ and $k$ $ >0$. These models describe a particle in an uniform circular motion with the frequency $\omega_{LM}=\frac{g}{m}$ and $\omega_{CO}=\frac{k}{g}$. A necessary condition to the LM/CO duality is to set $g^{2}=km$ so $\omega_{LM}=\omega_{CO}=\omega$.

As usual, duality means that both models predict the same effects in an indistinguishable form, i.e., there is no possibility of discriminating them by means of any kind of experiment. In the classical vision, it is only necessary that both models supply the same equations of motion because they give the only observables that we can consider (which are related with their solutions). Duality always must be related with all the observables that we can consider in each situation. In the LM/CO case, their classical solutions are the same if we do not consider a trivial solution, a constant vector
that appears in the LM reflecting the fact of this theory is invariant by a translation of the orbit's center.

The presence of duality in these mechanical models is completely natural since they can be seen as the infrared limit of the Maxwell-Chern-Simons (MCS) and Self-Dual (SD) field theories, which are two different ways to describe, in (2+1) dimensions, the dynamics of a single, freely propagating spin one massive mode with well defined helicity \cite{mcs-sd}. This MCS/SD duality can be demonstrated by several forms such as master action \cite{DJ}, Buscher procedure, canonical transformations or through the recently introduced Noether gauge embedding technique \cite{buscherlozanokquadrado}. An important feature of the duality is the presence of an explicit symmetry in one model that is hidden in the other.  In the MCS gauge symmetry is manifest while the SD model lacks that symmetry.  Similarly the LM is invariant under the translation of the orbit's center while the CO is not.

The LM/CO classical duality can be proved by the following master Lagrangian
\be
\label{mestra}
L_{M} = \frac{g}{2}\epsilon_{ij}{y}_{i}\dot{y}_{j}+g\epsilon_{ij}{x}_{i}\dot{y}_{j}+\frac{k}{2}x_{i}x_{i},
\ee
following the same arguments used in \cite{DJ} to demonstrate the MCS/SD equivalence.
It also follows through the dual projection approach \cite{dualprojection} which is operational for us here.  This is done by reducing the LM (\ref{LM}) to its first order form 
followed by a canonical transformation $\{x_i,p_i\}\to \{x_i^+ , x_i^-\}$  as $x_i = (x_i^+ + x_i^-)$ and $2p_i = g \epsilon_{ij} (x_j^- -x_j^+)$,   
that diagonalizes the LM model as $L_{LM}(\vec{x},\vec{p}) \to L_{CO}^{(+)}(\vec{x}^{+}) + L_{CS}^{(-)}(\vec{x}^{-})$ where $L_{CO}^{(+)}(\vec{z})=+\frac{g}{2}\epsilon_{ij}z_{i}\dot{z}_{j}-\frac{k}{2}z_{i}z_{i}$ and $L_{CS}^{(-)}(\vec{z})=-\frac{g}{2}\epsilon_{ij}z_{i}\dot{z}_{j}$. The last component represents the pure ``mechanical CS" Lagrangian which produces an additional constant vector representing the other solution of the LM. In the classical context, the LM/CO duality is totally transparent because $L_{CS}^{(-)}$ is physically sterile -- it carries a representation of the translation invariance of the LM. This term, without any dynamics, is a pure manifestation of this symmetry. Therefore, the LM action can be seen as a sum of two clearly separate pieces: one manifesting the dynamics ($L_{CO}^{(+)}$) and the other carrying the symmetry ($L_{CS}^{(-)}$). 

Now, consider the quantum version of $L_{CO}^{(+)}$. From the commutator $[x^{+}_{i}, x^{+}_{j}]=-\frac{i\hbar}{g}\epsilon_{ij}$, determined by its symplectic structure \cite{mqchernsimons, fadjack}, and Hamiltonian $H^{+}= \frac{k}{2}x^{+}_{i}x^{+}_{i}$, one recognizes that this model has the structure of an one-dimensional harmonic oscillator \cite{commentA}. It should be noticed that the coordinates of the CO system are noncommutative like the coordinates of the electrons in the lowest Landau level \cite{girvin}. One can see, by means of the Noether theorem, that the angular momentum operator $M^{+}$ and the Hamiltonian $H^{+}$ are proportional and so their spectrum
is
\be
\ M^{+}|n^{+}\rangle = \frac{H^{+}}{\omega} |n^{+}\rangle= \hbar \biggl(n^{+} + \frac{1}{2}\biggr)|n^{+}\rangle,   
\label{spectrumhamiltonianchiral}
\ee 
where $n^{+}=0,1,\ldots$. Thus, the angular momentum spectrum consists of positive half-integers. Similarly, the $L_{CS}^{(-)}$ has an angular momentum operator $M^{-}$ with the spectrum $M^{-}|n^{-}\rangle= -\hbar \biggl(n^{-} + \frac{1}{2}\biggr)|n^{-}\rangle$, with $n^{-}=0,1,\ldots$. The $L_{CS}^{(-)}$ describes a CO in the limit of vanishing energy, i.e., $H^{-}=0$. As so it is responsible for the degeneracy of the LM \cite{commentB}. All the energy (the dynamics) of the LM is contained in its internal CO, in agreement with the classical picture. 

The uncommon behavior of the CO system through rotations should be expected because it is, effectively, an one-dimensional system. Having this in mind, it is fruitful to consider more deeply the invariance of the considered systems through rotations. It is clear that both the LM and the CO are invariant under ``global" rotations of the kind $\delta x_{i}(t)=-\lambda \epsilon_{ij}x_{j}(t)$, $\lambda$ being time independent, and that their respective angular momentum are the generators of these transformations. To promote this ``global" symmetry to a ``local" gauge symmetry with time dependent $\lambda$, we follow \cite{mqchernsimons, maslov} and introduce a ``gauge potential" $a(t)$ which produces the covariant time derivative $Dx_i \equiv \dot{x_i} + a\epsilon_{ij}x_j$ and the local rotation $\delta x_{i}(t)=-\lambda(t)\epsilon_{ij}x_{j}(t)$, where $\delta a(t)=\dot{\lambda}$. 
In this dimensionality, only a further true CS term can be added to the Lagrangian and it must be linear in $a$. The generalization of (\ref{CO}) would be
$L_{CO}^{\nu}= \frac{g}{2}\epsilon_{ij}x_{i}D{x}_{j} + \nu a$, and for the LM (\ref{LM}) we would have $L_{LM}^{\nu}= \frac{m}{2}D{x}_{i}D{x}_{i} - \frac{g}{2}\epsilon_{ij}x_{i}D{x}_{j} + \nu a$.
This $\nu$ parameter is identified with the angular momentum in each model, respectively. 

Under a gauge transformation, the CS term in the action $\nu \int a dt $ transforms as $\nu \Delta \lambda$, where $\Delta \lambda$ is the change of the gauge function $\Delta \lambda=\int dt (d/dt)\lambda$. These gauge transformations are classified by requiring $\Delta \lambda=2\pi \mathcal{N}$, where $\mathcal{N}$ is the winding number of this $U(1)$ gauge transformation. Since $\Pi_{1}(U(1))=Z$, we could expect that in both quantum theories discussed $\nu \in Z$, and thus the angular momentum would be generically like $M_{z}=\nu \hbar$. However, a quantum anomaly changes this situation for the CO. Using the complex coordinates $z=x+iy$ and $z^{\dag}=x-iy$, the Lagrangian $L_{CO}^{\nu}$ is written (apart from a total time derivative) as $L_{CO}^{\nu}= -\frac{g}{2}z^{\dag}\left(i\frac{d}{dt} + a\right)z + \nu a$ and its effective quantum action is $\Gamma_{CO}^{\nu}(a)= -i$ ln det$\left(i\frac{d}{dt} + a\right) +\nu \int a  dt$. The determinant is not invariant against gauge transformations with nontrivial winding number \cite{determinant, mqchernsimons, maslov} because under this ``large" gauge transformation it acquires the sign $(-1)^{\mathcal{N}}$, what is usually called a $Z_2$-anomaly \cite{wittenanomaly}. Hence, to ensure the mathematical consistency of the $L_{CO}^{\nu}$, a ``large" ($\mathcal{N}$$ \ne 0$) gauge invariance of $e^{\frac{i}{\hbar}\Gamma_{CO}^{\nu}(a)}$ requires the half-integer quantization of $\nu$. This is the effect of the anomaly in the CO. On the other hand, the LM response to the anomaly is quite different. The $L_{LM}^{\nu}$ is a product of two anomalous determinants (the $(-1)^{\mathcal{N}}$ of the symmetry sector CS cancels the $(-1)^{\mathcal{N}}$ of the dynamical sector CO), and therefore $\nu \in Z$ and $M_{z}=\nu \hbar$. The LM is anomalous free but the duality was broken because the angular momentum of the LM always differ of the CO by a $\frac{\hbar}{2}$. 

Now, we discuss the theoretical basis of an idealized experimental setup where this difference could be measured. While experimental arrangements for the LM are quite common in the quantum Hall effect, the CO still does not have an experimental setup well developed. However, a few years ago, Baxter \cite{baxter} showed that by a suitable experimental arrangement the gross motion of a cold Rydberg atom with dipole's moment $\vec{d}$ could have a Lagrangian description by means of a CO. We adopt this construction as a possible manner to produce experimentally a CO system, and now we review some important points addressed there. Following \cite{baxter}, the Lagrangian
\be
\ L_{0}= \frac{m\dot{\vec{x}}^{2}}{2}-\dot{\vec{x}}.(\vec{d} \times \vec{B}(\vec{x})) + \vec{d}.\vec{E}(\vec{x})
\label{dipole}
\ee    
describes, in three dimensional space, a dipole (cold Rydberg atom) of moment $\vec{d}$ and mass $m$, in a presence of eletric $\vec{E}$ and magnetic $\vec{B}$ fields. The middle term in this expression is the R\"ontgen energy necessary to conserve momentum and guarantee gauge invariance \cite{baxter,rydbergtheory}. The motion of the dipole is restricted to two dimensions by the application of eletric and magnetic fields leading the R\"ontgen interaction to take a ``mechanical" Chern-Simons (CS) appearance, i.e., $\dot{\vec{x}}.(\vec{d} \times \vec{B}(\vec{x}))=\frac{g}{2}\epsilon_{ij}\dot{x}_{i}x_{j}$ and $\vec{d}.\vec{E}(\vec{x})=-\frac{k}{2}x_{i}x_{i}$. In this restricted system, (\ref{dipole}) can be written as
\be
\ L_{1}= \frac{m\dot{x}_{i}\dot{x}_{i}}{2} + \frac{g}{2}\epsilon_{ij}{x}_{i}\dot{x}_{j}-\frac{k}{2}x_{i}x_{i}.
\label{dipolo2d}
\ee
The parameter $g$ is proportional to the magnitude of the dipole moment and also dependent on the magnetic field. The harmonic term ($k>0$) in the above expression can be constructed by an optical trapping field. The possibility to produce a CO system arises when the speed of the atom is decreased \cite{trapping} in a such way that the kinetic energy term above may become irrelevant, transforming (\ref{dipolo2d}) in (\ref{CO}) \cite{baxter}. In \cite{mqchernsimons} the model (\ref{dipolo2d}) was used to study the reductibility properties just mentioned. 

Now, consider that one has a LM and a CO in hands. Is it possible to differentiate them? As it was previously told, the duality is broken at quantum level and therefore the answer must to be yes. In fact it is only necessary to detect the presence of the $\frac{\hbar}{2}$ in the angular momentum spectrum of the CO system. To measure this effect, an approach similar to te one introduced in \cite{interferometry} can be used: consider a conventional Mach-Zender interferometer in which the beam pair spans a two dimensional Hilbert space $\tilde{\mathcal{H}}=\lbrace |\tilde{0}\rangle, |\tilde{1}\rangle \rbrace$. Mirrors, beam splitters, and relative $U(1)$ phase shifts are represented by the unitary operators $\tilde{U}_{M}$, $\tilde{U}_{B}$ and $\tilde{U}(1)$ respectively, as in \cite{interferometry}.    
Sj\"oqvist {\it et al.} considered the case where the input particles have other degrees of freedom like spin. In our case these other degrees of freedom are represented by a CO in its ground state $|0\rangle$, with the associated density operator $\rho_{0}=|0\rangle \langle 0|$. This operator could be made to change inside the interferometer $\rho_{0} \to U_{i}\rho_{0}U_{i}^{\dag}$, with $U_{i}$ being a unitary transformation acting only on the CO system. Mirrors and beam splitters are assumed to leave the internal state unchanged so that we may replace $\tilde{U}_{M}$ and $\tilde{U}_{B}$ by $U_{M}=\tilde{U}_{M} \otimes 1_{i}$ and $U_{B}=\tilde{U}_{B} \otimes 1_{i}$, respectively, $1_{i}$ being the internal unit operator. Furthermore, one introduces the unitary transformation $\ U = \left( \begin{array}{ccc}
0 & 0 \\
0 & 1 \\
\end{array} \right) \otimes U_{i} + \left( \begin{array}{ccc}
e^{i\chi} & 0 \\
0 & 0 \\
\end{array} \right) \otimes 1_{i}$, where $U$ corresponds to the application of $U_{i}$ along the $|\tilde{1}\rangle$ path and the $U(1)$ phase $\chi$ similarly along $|\tilde{0}\rangle$. 

Now, let $\tilde{\rho}_{in}=|\tilde{0}\rangle \langle\tilde{0}|$ and the total incoming state given by the density operator $\rho_{in}=\tilde{\rho}_{in} \otimes \rho_{0}$. It is split coherently by a beam splitter and recombined at a second beam splitter after being reflected by two mirrors. The $U$ operator is applied between the first beam splitter and the mirror pair. Thus, the incoming state transforms into the output state $\rho_{out}=U_{B}U_{M}UU_{B}\rho_{in}U_{B}^{\dag}U^{\dag}U_{M}^{\dag}U_{B}^{\dag}$ and using the definitions above cited, we would have

\ba
\ \rho_{out}&=&\frac{1}{4}\left[\left( \begin{array}{ccc}
1 & 1 \\
1 & 1 \\
\end{array} \right) \otimes U_{i}\rho_{0}U_{i}^{\dag} + \left( \begin{array}{ccc}
1 & -1 \\
-1 & 1 \\
\end{array} \right) \otimes \rho_{0} \right. \nonumber\\
&+& \left. e^{i\chi}\left( \begin{array}{ccc}
1 & 1 \\
-1 & -1 \\
\end{array} \right) \otimes \rho_{0}U_{i}^{\dag} \right. \nonumber\\
&+& \left. e^{-i\chi}\left( \begin{array}{ccc}
1 & -1 \\
1 & -1 \\
\end{array} \right) \otimes U_{i}\rho_{0}\biggr] \right.
\label{manymatrices}
\ea
and therefore, the output intensity along $|\tilde{0}\rangle$ is
\be
\ I 
\propto 1 + |Tr\left(U_{i}\rho_{0}\right)|cos\left[\chi - arg Tr \left(U_{i}\rho_{0}\right)\right],
\label{intensity}
\ee
where $Tr\left(\rho_{0}U_{i}^{\dag} \right)=\left[Tr\left( U_{i}\rho_{0}\right)\right]^{*}$. 
The phase $\chi$ is shifted by $\xi = arg Tr\left( U_{i}\rho_{0}\right)$ what, in our case, corresponds to the Pancharatnam phase difference between $U_{i}|0 \rangle$ and $|0 \rangle$, \cite{pancharatnam}. If we set $U_{i}=e^{\frac{i}{\hbar}\alpha M_{z}}$ and considering $\alpha=2\pi$ rotations, we would have $\xi = \pi$, as in a spinorial system. Detect this $\pi$-phase shift is to measure the presence of the $Z_{2}$-anomaly in the CO and consequently, the break of LM/CO quantum duality.

The destructive role performed by the $Z_{2}$-anomaly discussed here in the LM/CO duality reveals the striking influence of anomalies over quantum dualities, that can have strong consequences to several dualities discussed nowadays \cite{futuro}. A situation, analogous to the LM/CO case, was found in a field theoretical model \cite{abreu} where the partition function for the Siegel formulation \cite{siegel} of 2D chiral boson was factorized in two contributions: the Floreanini-Jackiw partition function \cite{fj}, carrying the chiral dynamics and the noton contribution \cite{hull}, carrying a representation of the Siegel symmetry, which acquires dynamics at the quantum level. Therefore, the classical equivalence between the Siegel and the Floreanini-Jackiw models \cite{Bernstein:1988zd} is broken at the quantum domain. These examples carries us to believe that similar phenomenon could be present in other systems \cite{futuro}. 

It is important to stress that experimental verifications of quantum dualities are rare. The fact that most of the dualities discussed nowadays are in the context of quantum field theory and that the current knowledge of these theories is limited to weak coupling region produce the misleading idea that it is impossible to test dualities in general. However, a few years ago a charge-flux duality was proposed in the context of the fractional quantum Hall effect \cite{zhang} and, subsequently, this duality was indeed observed \cite{dualityHall}. 
The fact that some of the features of the LM/CO equivalence introduced here, as for example the destruction of quantum duality by effects created by a $Z_{2}$-anomaly, could in principle be verified by means of the current technology involving the manipulation of cold Rydberg atoms is, therefore, fascinating.

{\bf Acknowledgments}: This work is partially supported by CNPq, FUJB, CAPES, and FAPERJ,
Brazilian scientific agencies.

{\it E-mail addresses}: noronha@cbpf.br (J.L. Noronha), clovis@if.ufrj.br (C. Wotzasek).

\ed

To The Editor,
Physics Letter B
Professor M. Cvetic
Dept. of Physics and Astronomy
University of Philadelphia
Philadelphia, PA 19106, USA
cvetic@cvetic.hep.upenn.edu

Dear Editor,

we are submitting our paper entitled "Testing quantum duality using cold Rydberg atoms" to be considered for
publication in the Physics Letter B.

                              Thanking you,
                            Yours sincerely,
                              C.Wotzasek
                       e-mail: clovis@if.ufrj.br